\documentclass[aps,prb,twocolumn,groupedaddress]{revtex4}

\usepackage{epsfig}
\usepackage{amsfonts}

\begin{document}

\title{Optical Resonances in Reflectivity near Crystal Modes with Spatial
  Dispersion}
\author{I.~Kaelin, Ch.~Helm, and G.~Blatter}
\affiliation{Institut f{\"u}r Theoretische Physik,  ETH H{\"o}nggerberg, 
8093 Z{\"u}rich, Switzerland}

\date{\today}

\begin{abstract}
We study the effect of spatial dispersion of crystal modes on optical
properties such as the reflectivity $R$.  
As an example for isotropic media, we investigate the simplest model for
phonons in ionic crystals  and compare with previous results for
highly anisotropic plasmons, which are now understood from a more general point of view.  
As a consequence of the wave vector dependence of the dielectric
function small changes in the lineshape
are predicted. Beyond that, if the frequency of minimal $R$ is near a pole
of the dispersionless dielectric function, the relative amplitude of dips in
$R$ with normal 
and anomalous dispersion differ significantly, if dissipation and
disorder are low. 
\end{abstract}

\maketitle

Usually, in crystal optics the dependence of the dielectric function 
on the wave vector $\mathbf k$ is
neglected because the wave length $\lambda$ is much larger than interatomic
length scales $s$ and effects of spatial dispersion are expected to be of the
order of $s/\lambda\ll 1$, see Refs.~\onlinecite{Gin, Hen} and references 
therein. 
Recently, it was shown for the reflectivity near the 
Josephson plasma resonance (JPR) in highly anisotropic superconductors
\cite{Hel02a,Hel02c}, which is an interlayer charge oscillation 
of Cooper pairs, that spatial dispersion provides a novel way to stop light
 and can have effects of order
unity on the relative amplitude of plasma modes with different dispersion. 
This system is exceptional with respect to the
symmetry (high anisotropy), the fact that light is screened away from the JPR, 
and the type of the excitation (plasmon). 
Here we discuss the possibility that such effects appear under 
more general conditions. As a generic example of a spatially dispersive
resonance we choose 
the simplest model of one optical phonon band in an ionic crystal
\cite{Mar} (also describing certain excitons \cite{Gin}) and study
the optical properties of the isotropic phonon-polariton, 
which is the lattice vibration coupled to light. 
Thereby, we determine under what circumstances similar effects can be expected 
in a large class of systems and clarify
the origin of previous results for the highly anisotropic plasmon obtained
in Ref.\ \onlinecite{Hel02a}.
We show that the effect of spatial dispersion is of order unity,
if the resonance in reflectivity is in a frequency region of size 
$\Delta\omega/\omega_0=s/\lambda$ near the 
pole $\omega_0$ of the dispersionless dielectric function $\epsilon(\omega)$; 
the amplitudes in reflectivity of phonon bands with normal and anomalous
dispersion then differ significantly. 

Weak disorder and small dissipation turn out to be essential for the
observation of significant effects due
to spatial  dispersion, which is the 
case in the plasmon system of the JPR below the superconducting gap. In
insulators, e.g.\ in ionic crystals with perfect lattice symmetry and weak
disorder, such favorable conditions might be realized as well, while
in metals similar effects would be strongly overdamped.

In the following, we mainly consider an electromagnetic wave of
frequency $\omega$ in normal incidence, with a wave vector $\mathbf
k=(0,0,k_z)$ at the surface $z=0$ of a semi-infinite dielectric medium. 
The macroscopic Maxwell equations describing the electromagnetic fields
inside the medium expressed in Fourier components   take the form 
\begin{equation}
\label{ME1}
k_zE_x  = (\omega/c)B_y, \qquad 
k_z B_y  =  (\omega/c) \epsilon(\omega,\mathbf k)E_x,
\end{equation}
where $E_x$ and $B_y$ are the components of the electric 
and the magnetic fields, respectively. 
For an isotropic medium the normal modes are determined by 
\begin{equation}
\label{DRISO}
\frac{k^2}{\epsilon(\omega,\mathbf k)}=\frac{\omega^2}{c^2}. 
\end{equation}
For definiteness,  we consider the spatially
dispersive dielectric constant 
$\epsilon(\omega,q)$, $q=sk_z$, $q\in [0,2\pi]$ ($s$ lattice constant), of an
ionic crystal, which can be obtained by solving the microscopic equation of
motion for the ions \cite{Gin} (using the force constants of the bulk),
\begin{equation}
\label{DK}
\epsilon(\omega,q)=\epsilon_{\infty}+\frac{S\omega_q^2}{\omega_q^2-\omega^2-i\rho\omega},
\end{equation}
where $\epsilon_{\infty}$ is the background dielectric constant, 
$S=S_0(1+\alpha q^2)$, $\alpha \sim 10^{-3}$, $S_0\sim 1$, 
is the oscillator strength, 
$\omega_q^2=\omega_0^2(1+dq^2)$, $\omega_0/(2\pi)\sim 1$ THz, $d\sim 0.1$, is the dispersion 
of the crystal mode, 
and $\rho$ is the damping due to
disorder ($\rho/\omega_0\sim 10^{-2}$ for NaCl, see Ref.~\onlinecite{BruI}). 
Since $\alpha/d\ll 1$, the wave vector dependence
of the oscillator strength can be neglected. The sign of $d$ determines
the shape of the dispersion and both $d>0$ (normal dispersion) and $d<0$
(anomalous dispersion) naturally appear in ionic crystals due
to the backfolding of the modes into the first Brillouin zone (e.g.\ realized
in fcc structured NaCl for the directions $\mathbf k \| [100]$ and $\mathbf k
\|[111]$, see Ref.~\onlinecite{Mar}). The dielectric function in
Eq.~(\ref{DK}) is also a model for certain excitons where the damping 
$\rho/\omega_0 \sim 10^{-4}$ is much smaller, see Ref. \onlinecite{Gin}.

The formula analogous to Eq.~(\ref{DRISO}) for an anisotropic plasmon such as
the JPR reads
\begin{equation}
\label{DRANISO}
\frac{k_x^2}{\epsilon_c(\omega,k_z)}+\frac{k_z^2}{\epsilon_a(\omega)}=\frac{\omega^2}{c^2},
\end{equation}
where the spatially dispersive dielectric function $\epsilon_c(\omega,
q)=\epsilon_{c0}(1-\omega_c^2(q)/\omega^2)$, $q=sk_z$, $\omega_c^2(
q)=\omega_{c0}^2(1+dq^2)$,
accounts for the $c$ axis plasmon (excitation parallel to the $z$ axis).
This mode is only excited for light in oblique incidence $\mathbf
k=(k_x,0,k_z)$, where 
$k_x=\omega\sin\theta/c$ is conserved due to translational invariance parallel to the
surface ($\theta$ is the angle of incidence).
The dielectric constant parallel to the surface
$\epsilon_a(\omega)=\epsilon_{a0}(1-\omega_a^2/\omega^2)$ is non-dispersive. 
The resonance frequency $\omega_a$ for the oscillation parallel to the surface
is far above the $c$ axis plasmon frequency, $\omega_a\gg \omega_{c0}$, and for
frequencies near $\omega_{c0}$ it follows that $\epsilon_a(\omega)\simeq
-\omega_a^2/\omega^2\ll -1$. 

\begin{figure}
\epsfig{file=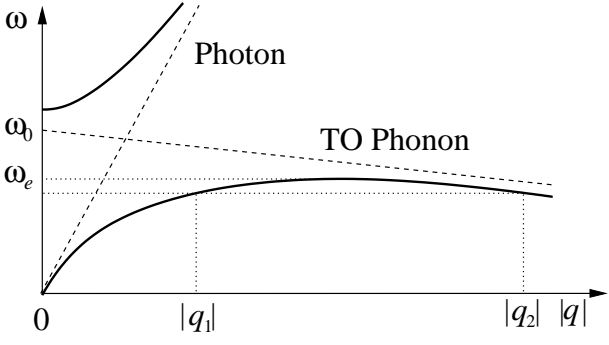,width=0.45\textwidth,clip=}
\caption{Polariton dispersion relation $\omega(|q|)$ for a transverse optical
(TO) phonon. Multiple modes $q_1, q_2$ are excited by incident light of
frequency $\omega$ according to the solutions of Eq.~(\ref{DRISO}). 
The appearance of an extremal frequency $\omega_e$, where the
group velocity $v_{gz}=\partial\omega/\partial k_z$ vanishes and $q$ changes sign, 
is characteristic for the mixing of normally dispersive light and a crystal
 mode with  anomalous dispersion. \label{polariton}}
\end{figure}
\begin{figure}
\epsfig{file=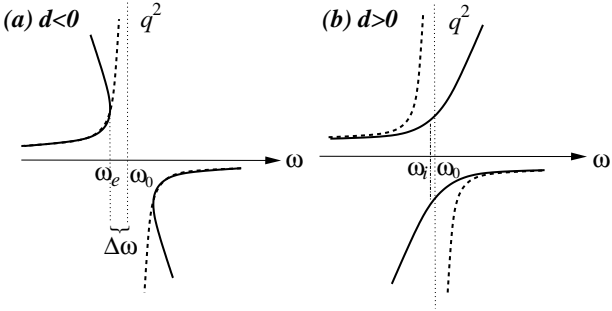,width=0.45\textwidth,clip=}
\caption{A pole $\omega_0$ in $k^2(\omega)$ in the case without
spatial dispersion (dashed lines) is regularized when spatial dispersion is taken
into account (solid), see also  the inverse $\omega(k^2)$ in
Fig.~\ref{der}$(a)$. Depending on the sign of $d\propto \partial
k^2(\omega)/\partial \omega$, an extremal point $\omega_e$, where the group
velocity $v_{gz}=\partial\omega/\partial k_z$ vanishes, occurs in case of anomalous
phonon dispersion $(a)$ and  an extremal point $\omega_i$, 
where $k_1^2(\omega_i)=-k_2^2(\omega_i)$, 
appears in case of normal phonon dispersion $(b)$. \label{reg}}
\end{figure}
Focusing on the phonon case first, the dispersion relation $\omega (q)$ (see
Fig.~\ref{polariton}) follows from Eq.~(\ref{DRISO}) with $\epsilon
(q,\omega)$ from Eq.~(\ref{DK}) and shows that at a given frequency $\omega$
two modes with (generally complex) refraction indices $n_p=c q_p/s\omega$, $p=1,2$,
can be excited due to spatial dispersion.
This corresponds to the regularization (cf. Fig.~\ref{reg}) of the pole in 
$k^2(\omega)\propto \epsilon(\omega)$, cf. Eq.~(\ref{DRISO}), which exists in
the dispersionless dielectric function $\epsilon(\omega)$ at the phonon
resonance $\omega_0$ in the absence of damping ($\rho=0$). 
Thereby the appearance of an extremal
frequency $\omega_e$, where the group velocity $v_{gz}=\partial
\omega/\partial k_z$  for propagation in the $z$ direction vanishes, 
is characteristic for the mixing of the normal dispersive light mode and
an anomalous crystal mode in Fig.\ \ref{reg}$(a)$ (cf. Fig.~\ref{polariton}), 
while  for equal dispersion of light and crystal modes a special 
frequency $\omega_i$ appears, where $q_1^2(\omega_i)=-q_2^2(\omega_i)$, see
Fig.~\ref{reg}$(b)$. Near $\omega_{e,i}$ the amplitude
and lineshape in the reflectivity can be sensitive to the type of dispersion,
see below.

In a semi-infinite crystal, the causality principle requires that  
$v_{gz}$ is positive, thus only  two of the four solutions of Eq.~(\ref{DRISO}) for
$n_p$ are physical, e.g., for an anomalous dispersion the real part of $n_p$ is negative. 
When $|n_1| \ll |n_2|$, only the small-$q$ mode
$n_1$ is light like and expected to affect optical properties, while for 
$|n_1| \approx |n_2|$ both modes have to be considered. 
Note that for two real modes near $\omega_e$, where $v_{gz}=0$ (Fig.~\ref{polariton}),
causality demands $n_1=-n_2$.

The excitation of multiple modes poses the 
problem that the usual Maxwell boundary conditions, requiring continuity of the
electric and magnetic fields parallel to the surface, 
are insufficient to calculate the relative amplitudes of these modes.
In the past, this problem was
addressed in a phenomenological way by introducing so called additional
boundary conditions (ABC) for the macroscopic polarization $\mathbf P$.  
We  use  the general ABC  \cite{Gin}
\begin{equation}
\label{ABCGIN}
\mathbf{P}(z)+l\left(\partial \mathbf P/\partial z\right)=0 \quad (z\to 0),
\end{equation}
where $l$ is a phenomenological  decay length
and 
$\mathbf P=\chi(\omega,\mathbf k)\mathbf E$, 
with $\chi(\omega,\mathbf k)=
\left(\epsilon(\omega,\mathbf k)-\epsilon_\infty \right)/4\pi$ for the isotropic
medium.
These ABC have been justified for the anisotropic plasmon in a microscopic model by
subtracting the true equation of motion for
the surface layer from the hypothetical bulk equation for this layer, 
see Ref.~\onlinecite{Hel02a}. 
The analogous analysis for the phonon allows to identify the parameter $l$ in 
Eq.~(\ref{ABCGIN})  with the nearest neighbor separation $s/2$ in fcc 
cubic crystals (e.g.\ NaCl).

\begin{figure*}
\epsfig{file=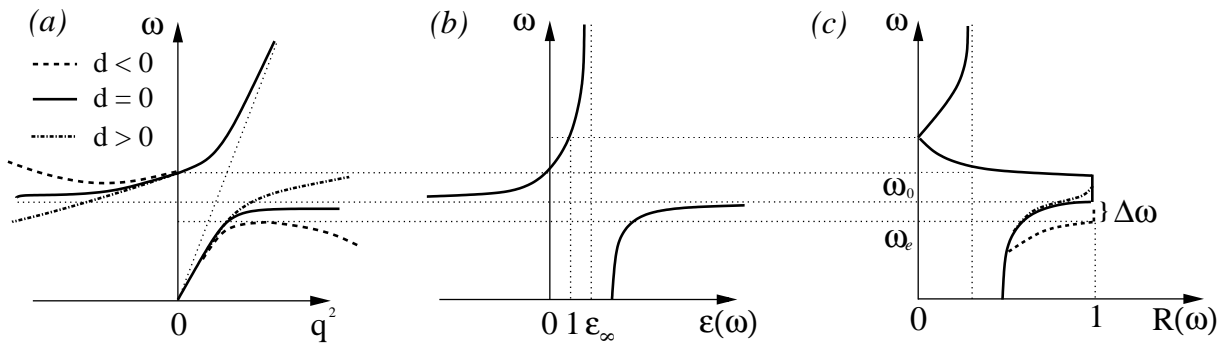,width=0.9\textwidth,clip=}
\caption{
  $(a)$ Polariton dispersion relation $\omega(q^2)$ for different crystal
  dispersion $d$, where $q^2>0$ ($q^2<0$) corresponds to propagating (decaying)
  modes. 
 $(b)$ Dielectric function $\epsilon(\omega)$ in the absence of
  spatial dispersion. $(c)$ Reflectivity $R(\omega)$ of the phonon band.
  In the region $\Delta\omega/\omega_0 = (\omega_0 - \omega_e) / \omega_0
  \sim s/\lambda$ near $\omega_0$ the lineshape is affected by spatial 
  dispersion, but the amplitude $1-R_{\rm min}$ of the dip away from this
  frequency region is not modified. 
\label{der}}
\end{figure*}

Given the boundary conditions on $\mathbf E$ and $\mathbf B$, we determine the
reflectivity $R$ at the boundary $z=0$ to the vacuum, which
is the ratio of the reflected and the incident energy flux density, 
\begin{equation}
\label{FRESNEL}
R=|r|^2, \quad r=\frac{1-\kappa}{1+\kappa}, \quad \kappa
=\frac{E_x(z=0)}{\cos\theta B_y(z=0)}.
\end{equation}
In the dispersionless case the reflectivity $R$ shows a dip with a minimal 
value $R_{\rm min}=0$ at $\epsilon(\omega) = \kappa^2 =1$, while the frequency
range of total reflection, $R=1$, between the pole and the zero of $\epsilon$ 
is bounded by sharp edges in $R(\omega)$, see Fig.~\ref{der}.
In the following we will show how spatial dispersion can reduce the
amplitude $1-R_{\rm min}$ of the dip in $R$ near $\kappa=1$, and modify the 
lineshape of $R$ close to the region of total reflection (cf. Fig.~\ref{der}).

For a spatially dispersive medium in the simplest case of normal incident
light, where two transverse modes $\mathbf q_p=(0,0,q_p)$, $p=1,2$, are
excited, we expand the electric field 
$E_x=\sum_{p=1}^{2}E_x(q_p)e^{iq_pz/s}$ and the polarization field 
$P_x=\sum_{p=1}^{2}E_x(q_p)\chi(\omega,\mathbf q_p)e^{iq_pz/s}$. 
Using Eqs.~(\ref{ME1}), (\ref{ABCGIN}), and (\ref{FRESNEL}) we
derive 
\begin{equation}
\label{KAPPAE}
\kappa_{\rm iso}=\frac{n_1n_2}{n_1+n_2-i\xi n_1n_2},
\end{equation}
in the isotropic system near the extremal frequencies $\omega_{e,i}$ 
(ABC for $P_x$) with the parameter $\xi=l\omega/c=s/2\lambda\sim 10^{-5}$. 
Similarly, for the anisotropic plasmon with the incident field ${\bf E}$
 in the $xz$-plane \cite{Hel02a} (ABC for $P_z$), we get
\begin{equation}
\label{kappaplasmon}
\kappa_{\rm aniso}=\frac{1}{\epsilon_a\cos\theta}\frac{n_1n_2}{n_1+n_2-i\xi
  n_1n_2}.
\end{equation}
The above expressions for $\kappa$ enter the reflectivity $R$, see
Eq.~(\ref{FRESNEL}). The difference in 
Eqs.~(\ref{KAPPAE}), (\ref{kappaplasmon}) arises from the anisotropy
$\epsilon_a\not=\epsilon_c$ in the plasmon system, which can suppress the
value of $\kappa$ for a strong anisotropy $|\epsilon_a| \gg 1$. 
Away from $\omega_{e,i}$, where $|n_1|\ll |n_2|$, Eq.~(\ref{KAPPAE}) reduces to
$\kappa_{\rm iso}=n_1$ and Eq.~(\ref{kappaplasmon}) to 
$\kappa_{\rm aniso}=n_1/(\epsilon_a\cos\theta)$, i.e.  
the usual Fresnel theory is recovered and 
the smaller (light like) refraction index determines the optical
properties.

\begin{figure}
\epsfig{file=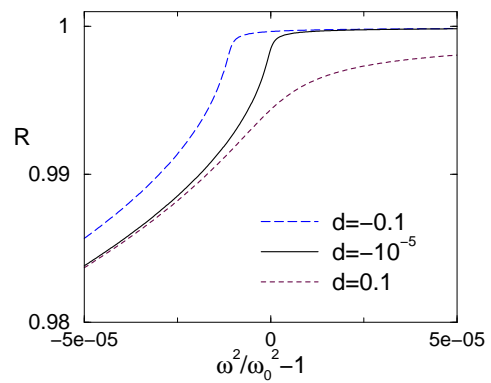,width=0.35\textwidth,clip=}
\caption{Lineshapes of the reflectivity $R$ near the upper band edge of the
  lower polariton branch for phonons with different dispersion $d$ in case of
  very low damping $\rho/\omega_0\sim 10^{-6}$ ($\epsilon_\infty =2$, $S_0=3$). 
  The shift of the band edge for anomalous phonon dispersion is of the order
  $\Delta\omega/\omega_0\sim\xi\sim 10^{-5}$ (distance between the
  solid and the dashed lines). In the region close to the phonon resonance
  $\omega_0$, where both modes $n_1, n_2$ are important, the lineshape 
  depends on the sign of $d$. \label{lineshape}}
\end{figure}

For $d<0$, in a frequency interval $\Delta\omega/\omega_0\sim\sqrt{|d|}s/\lambda
\sim \xi$ below $\omega_e$, where $n_1 \approx  -n_2$ and the group velocity
$v_{gz}$ vanishes, 
the standard Fresnel theory is invalid and small length scales,
which enter the reflectivity $R$ via the parameter $\xi$, are
relevant.  Note that at  $\omega_e$, where the (positive) solutions $q_p^2$ of
Eq.~(\ref{DRISO}) merge (cf.\ Fig.\ \ref{polariton}) and $n_1=-n_2$ due to
causality, $\kappa_{\rm iso}$ in Eq.~(\ref{KAPPAE}) is purely
imaginary and thus $R=1$, see Eq.~(\ref{FRESNEL}). Slightly away from
$\omega_e$, the amplitudes $1-R_{\rm min}$ of the dip
for isotropic phonons and anisotropic plasmons  equal to $(1+\xi^2)^{-1}\approx 1$ and 
$2/\left((1+\epsilon_a^2\xi^2\cos^2\theta)^{1/2}+1\right)<1$, respectively. 
As $\epsilon_a\xi\sim O(1)$,  $1-R_{\rm min}$ is
significantly suppressed in the anisotropic  system, while $\xi \ll 1$  is negligible for
the isotropic one. In addition, the
band edge near $\omega_e$ is slightly shifted to lower frequencies in
comparison with the dispersionless theory,
$(\omega_0-\omega_e)/\omega_0\sim \xi$ (Figs.~\ref{der}, \ref{lineshape}).

For $d>0$ near $\omega_i$ and above $\omega_0$, 
one solution $k^2$ of Eq.~(\ref{DRISO}) is positive, Fig.~\ref{der}$(a)$,
and thus creates a tail, where $R<1$ instead of $R=1$, see Fig.~\ref{lineshape}. 
The difference in reflectivity between normal and
anomalous phonon dispersion for damping  $\rho/\omega_0\sim \xi$ is
$\sim 1\%$ near the band edge $\omega_0$, see  Fig.~\ref{der}(a).

\begin{figure}
\epsfig{file=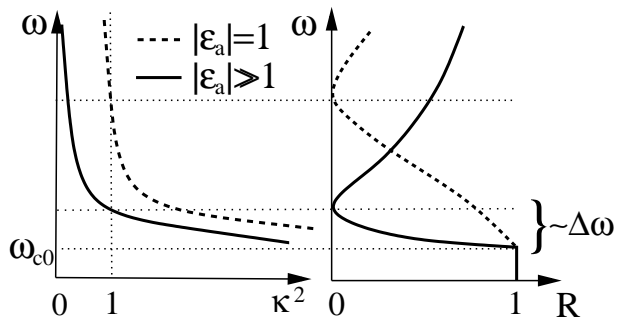,width=0.45\textwidth,clip=}
\caption{Reflectivity $R$ of the plasma band in the
  highly anisotropic system of Ref.~\onlinecite{Hel02a} (schematic). 
The minimum $R_{\rm min}$ near $\kappa = 1$ is
  in a frequency region $\Delta\omega/\omega_0 = (\omega - \omega_e) \sim \xi$ 
  close to the bandedge due to the high anisotropy $|\epsilon_a|\gg 1$ and
  spatial dispersion {\em can}  affect the amplitude of the dip $1-R_{\rm
  min}$.
 \label{plasmon}}
\end{figure}

These effects on the phonon lineshape in the reflectivity $R$ are small and, in
particular, the amplitude $1-R_{\rm min}$ of the dip above $\omega_0$ is hardly affected,
see Fig.~\ref{der}(c). In comparison with the anisotropic plasmon, where a
strong dependence of the dip amplitude on spatial dispersion was reported \cite{Hel02a}, we
seek to identify the crucial difference to the present system and under what circumstances such
an effect might appear in the phonon system. 

\begin{figure}
\epsfig{file=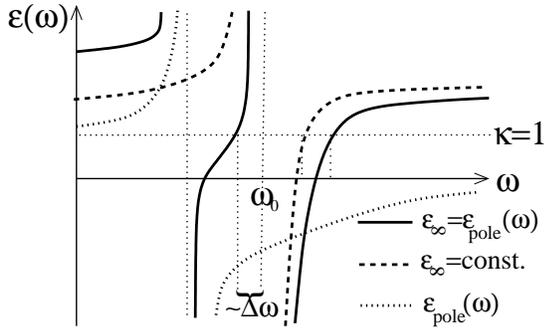,width=0.4\textwidth,clip=}
\caption{Dielectric function $\epsilon(\omega)$ with usual background 
dielectric
  constant $\epsilon_\infty \sim 1$ and in the presence of a background mode
  $\epsilon_{\infty} = \epsilon_{\rm pole} (\omega)$. In the latter case the
  minimum $R_{\rm min}$ at $\kappa=\sqrt{\epsilon}=1$ is shifted to the critical
  region of size
  $\Delta\omega/\omega_0 = (\omega_{e} - \omega_0) / \omega_0 \sim \xi$ around
  the pole $\omega_0$ (cf. Fig.~\ref{reg}), where spatial dispersion  can affect
  the optical properties.  \label{omtm}}
\end{figure}

Firstly, we conclude from Eq.~(\ref{DRANISO}) that in an anisotropic system
 ($\epsilon_a \neq \epsilon_c$) a pole in 
$q^2 \sim n^2 = \epsilon_a - (\epsilon_a/\epsilon_c) \sin^2 \theta$ like in
 Fig.~\ref{reg} is created  in oblique incidence ($\theta \neq 0$) 
at the zero of $\epsilon_c$ in a 
similar way as it appears in the isotropic case at the pole of 
$\epsilon (\omega)$. 
More importantly, the {\em strong} anisotropy $|\epsilon_a| \gg 1$ ($\sim 10^4$
 for JPR) of the order $\sim 1/\xi$ suppresses the value of $\kappa_{\rm aniso}$ in
 Eq.~(\ref{kappaplasmon}), such that the dip in $R$ near $\kappa=1$ is close to
 the bandedge and thus in the frequency range $\Delta \omega / \omega_0 \sim
 \xi$  near the special points $\omega_{e,i}$ (Fig.~\ref{reg}), where spatial
 dispersion is relevant (Fig.~\ref{plasmon}).

\begin{figure}
\epsfig{file=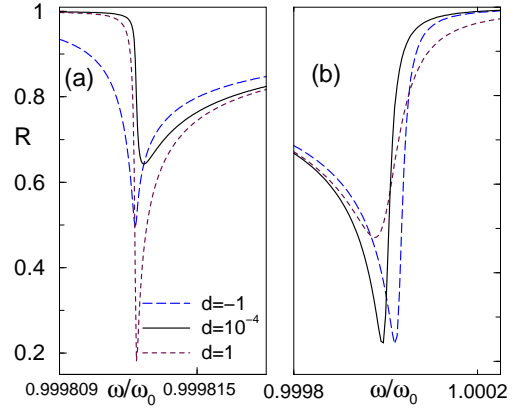,width=0.37\textwidth,clip=}
\caption{Reflection amplitude $R$ for a phonon with different
  dispersion $d$ and small dissipation (a)
  in the presence of a dispersionless background mode
 $\epsilon_\infty \rightarrow
  \epsilon_{\rm pole}$ ($S_b=1$, $\epsilon_\infty=3$, 
$\rho/\omega_0 \sim \rho_b / \omega_0 \sim  10^{-7} \ll \xi$) 
near the band edge at $\omega_b = 0.9995 \omega_0$,
  or (b) for almost parallel incidence 
($\theta \approx \pi / 2 - \xi^{1/2}$, $\rho/\omega_0 \sim  \xi$). In (b) for 
dispersion $d<0$ the  minimal reflectivity 
  $R_{\rm min} \simeq 0$, but the minimum is slightly shifted compared to the
  dispersionless model ($d\sim0$), while for $d>0$ the dip amplitude
  $1-R_{\rm min}$ is   strongly suppressed.
 \label{plotoftheday}}
\end{figure}

Guided by this observation, we find significant effects on the dip
amplitude $1-R_{\rm min}$ for the isotropic phonon, if the minimum in $R$ is
within the frequency interval $\Delta \omega / \omega_0 \sim s/\lambda$
near the pole of $\epsilon (\omega)$. This can be achieved (a) by a
second background dispersionless phonon,  $\epsilon_\infty \rightarrow \epsilon_{\rm pole}
(\omega) = \epsilon_\infty + S_b^{\phantom 2} \omega_b^2 / (\omega^2 - \omega_b^2 - i \rho_b
\omega)$, which is sufficiently close to $\omega_0$, $( \omega_0 - \omega_b)/
\omega_0 \sim s/\lambda \sim \xi$, i.e. $|\epsilon_{\rm pole} (\omega_0)| \sim
1/\xi \gg 1$ and creates an additional dip in $R$, see Fig.~\ref{omtm}, 
or similarly (b) in extreme oblique incidence ($\theta \approx \pi / 2 -\xi^{1/2}$), 
where the dispersionless $\kappa =  n / \epsilon \cos \theta = (\epsilon
-\sin^2 \theta)^{(1/2)} / \epsilon \cos \theta $ is $1$ at
$\omega_0 + O (\xi)$. This results in  significant differences in $R_{\rm min}$
for normal and anomalous dispersive phonon bands, if the dissipation is 
small, see Fig.~\ref{plotoftheday}. 
Physically the larger suppression of $1-R_{\rm min}$  for
dispersion $d>1$ is related to the fact that near $\omega_i$ only a
single propagating mode is excited, while the other one decays  into the crystal.
This effect is of order unity and the dip of width 
$\Delta\omega/\omega_0\sim \xi$ in $R$
can be resolved in optical phonon spectra \cite{Leo}.

In conclusion, for isotropic phonons small effects for the lineshape in
reflectivity due to spatial dispersion are predicted
(Fig.~\ref{lineshape}). For low dissipation
the dip amplitude  can depend significantly on the sign of the
dispersion, if the frequency of minimal reflection is in a small distance 
$\Delta \omega /\omega_0\sim s/\lambda$ to the dispersive resonance
(Fig.~\ref{plotoftheday}). This can be realized in strongly anisotropic
systems such as the JPR \cite{Hel02a} or in the isotropic case due to a second almost
degenerate resonance or for nearly parallel incidence.

We thank I. Biaggio, L. N. Bulaevskii, L. Degiorgi and
G. Montemezzani for useful discussions, and the Swiss
NSF in the NCCR ''MaNEP'' for financial support.

\end{document}